\documentstyle[12pt]{article}
\input amssym.def
\input amssym.tex

\newcommand{\wx}{\widetilde{x}}
\newcommand{\wip}{\widetilde{p}}
\newcommand{\wa}{\widetilde{a}}
\newcommand{\wb}{\widetilde{b}}
\newcommand{\wc}{\widetilde{c}}
\newcommand{\wid}{\widetilde{d}}

\newcommand{\qed}{\rule{3mm}{3mm}}

\begin{document}

\def\m@th{\mathsurround=0pt}
\mathchardef\bracell="0365 
\def\upbrall{$\m@th\bracell$}
\def\undertilde#1{\mathop{\vtop{\ialign{##\crcr
    $\hfil\displaystyle{#1}\hfil$\crcr
     \noalign
     {\kern1.5pt\nointerlineskip}
     \upbrall\crcr\noalign{\kern1pt
   }}}}\limits}

\begin{titlepage}{\LARGE
\begin{center} A new integrable system \\
related to the Toda lattice
 \end{center}} \vspace{1.5cm}
\begin{flushleft}{\large Yuri B. SURIS}\end{flushleft} \vspace{1.0cm}
Centre for Complex Systems and Visualization, University of Bremen,\\
Kitzb\"uhler Str. 2, 28359 Bremen, Germany\\
e-mail: suris @ cevis.uni-bremen.de

\vspace{1.5cm}  
{\small {\bf Abstract.} A new integrable lattice system is introduced,
and its integrable discretizations are obtained. A B\"acklund transformation
between this new system and the Toda lattice, as well as between their
discretizations, is established.}
 
\end{titlepage}

\setcounter{equation}{0}
\section{Introduction}

We want to introduce in this paper a new integrable lattice system:
\begin{equation}\label{new}
\ddot{x}_k=\dot{x}_k\Big(\exp(x_{k+1}-x_k)-\exp(x_k-x_{k-1})\Big),
\end{equation}
along with two integrable discretizations thereof. In the difference equations 
below $x_k=x_k(t)$ are supposed to be functions of the discrete time 
$t\in h{\Bbb Z}$, and $\wx_k=x_k(t+h)$, $\undertilde{x_k}=x_k(t-h)$. The first 
of our integrable discretizations is implicit with respect to the updates 
$\wx_k$'s:
\begin{equation}\label{dnew1}
\frac{\exp(\wx_k-x_k)-1}{\exp(x_k-\undertilde{x_k})-1}=
\frac{1+h\exp(\undertilde{x_{k+1}}-x_k)}{1+h\exp(x_k-\wx_{k-1})},
\end{equation}
and the other is explicit:
\begin{equation}\label{dnew2}
\frac{\exp(\wx_k-x_k)-1}{\exp(x_k-\undertilde{x_k})-1}=
\frac{1+h\exp(x_{k+1}-x_k)}{1+h\exp(x_k-x_{k-1})}.
\end{equation}

The system (\ref{new}) resembles much the usual Toda lattice:
\begin{equation}\label{Toda}
\ddot{x}_k=\exp(x_{k+1}-x_k)-\exp(x_k-x_{k-1}),
\end{equation}
and in fact turns out to be closely related to it my means of a sort of 
B\"acklund transformation. To the author's knowledge, this system has not 
appeared in the literature, despite its beauty and possible physical 
applications. 

It is by now well known that the Toda lattice admits several apparently different
integrable discretizations (which are, in fact, closely connected with each
other, but these connections are rather nontrivial). Two of the discretizations
are explicit with respect to $\wx_k$'s, namely the Hirota's one \cite{H}:
\begin{equation}\label{Hir}
\exp(\wx_k-x_k)-\exp(x_k-\undertilde{x_k})=
h^2\Big(\exp(x_{k+1}-x_k)-\exp(x_k-x_{k-1})\Big),
\end{equation}
and a standard--like one \cite{S1}:
\begin{equation}\label{stand}
\exp(\wx_k-2x_k+\undertilde{x_k})=\frac{1+h^2\exp(x_{k+1}-x_k)}
{1+h^2\exp(x_k-x_{k-1})},
\end{equation}
which can be also presented in the form
\begin{equation}\label{Sur1}
\exp(\wx_k-x_k)-\exp(x_k-\undertilde{x_k})=
h^2\Big(\exp(x_{k+1}-\undertilde{x_k})-\exp(\wx_k-x_{k-1})\Big).
\end{equation}
Another two discretizations are implicit with respect to $\wx_k$'s. They
were introduced in \cite{S3} and read:
\begin{equation}\label{Sur2}
\exp(\wx_k-x_k)-\exp(x_k-\undertilde{x_k})=
h^2\Big(\exp(\undertilde{x_{k+1}}-x_k)-\exp(x_k-\wx_{k-1})\Big)
\end{equation}
and
\begin{equation}\label{Sur3}
\exp(\wx_k-x_k)-\exp(x_k-\undertilde{x_k})=
h^2\Big(\exp(\undertilde{x_{k+1}}-\undertilde{x_k})-\exp(\wx_k-\wx_{k-1})\Big).
\end{equation}

We discuss their algebraic structure, as well as their relations to each
other further on.

We shall demonstrate that (\ref{dnew1}) is related to (\ref{new}) just in
the same way as (\ref{Sur2}) is related to (\ref{Toda}), and shall also 
elaborate an algebraic structure of the discretization (\ref{dnew2}).

All the systems above (continuous and discrete time ones) may be considered
either on an infinite lattice ($k\in{\Bbb Z}$), or on a finite one 
($1\le k\le N$). In the last case one of the two types of boundary conditions 
may be imposed: open--end ($x_0=\infty$, $x_{N+1}=-\infty$) or periodic
($x_0\equiv x_N$, $x_{N+1}\equiv x_1$). We shall be concerned only with the
finite lattices here, consideration of the infinite ones being to a large
extent similar.

\setcounter{equation}{0}
\section{Newtonian equations of motion:\newline
Lagrangian and Hamiltonian formulations}

All the equations introduced in the previous section, both continuous--
and discrete--time, are written in the Newtonian form:
\[
\ddot{x}_k=\Phi_k(\dot{x},x)\quad{\rm or}\quad \Psi_k(\wx,x,\undertilde{x})=0,
\]
respectively. They all turn out to admit a Lagrangian formulation.

Recall that in the continuous time case Lagrangian equations are given by
\begin{equation}\label{Lagr}
\frac{d}{dt}\frac{\partial{\cal L}}{\partial\dot{x}_k}-
\frac{\partial{\cal L}}{\partial x_k}=0,
\end{equation}
while their discrete time analog is given by
\begin{equation}\label{d Lagr}
\partial\Big(\Lambda(\wx,x)+\Lambda(x,\undertilde{x})\Big)/\partial x_k=0.
\end{equation}

Finally, recall that Lagrangian formulation implies also a possibility of
introducing a Hamiltonian one. Namely, in the continuous time case one
defines momenta $p_k$ canonically conjugated to the coordinates $x_k$ by
\begin{equation}
p_k=\partial{\cal L}/\partial\dot{x}_k.
\end{equation}
Then the flow defined by (\ref{Lagr}), being expressed in terms of $(x,p)$, 
preserves the standard symplectic form $\sum dx_k\wedge dp_k$ on the phase space
${\Bbb R}^{2N}(x,p)$.
Moreover, this flow may be written in a canonical form
\begin{equation}\label{Ham}
\dot{x}_k=\partial H/\partial p_k,\quad \dot{p}_k=-\partial H/\partial x_k,
\end{equation}
the Hamiltonian function $H(x,p)$ being given by
\begin{equation}\label{L to H}
H=\sum_{k=1}^N \dot{x}_kp_k-{\cal L}.
\end{equation}

Analogously, in the discrete time case the momenta $p_k$ canonically conjugated
to $x_k$ are given by
\begin{equation}\label{dLagr:x to p}
p_k=\partial\Lambda(x,\undertilde{x})/\partial x_k.
\end{equation}
Then the map $(x,\undertilde{x})\mapsto(\wx,x)$ induces a symplectic map
$(x,p)\mapsto(\wx,\wip)$ of the phase space ${\Bbb R}^{2N}(x,p)$, i.e. a map
preserving the standard symplectic form $\sum dx_k\wedge dp_k$. Note that
(\ref{dLagr:x to p}) implies that the equations (\ref{d Lagr}) may be 
presented as 
\begin{equation}\label{dLagr:p}
p_k=-\partial\Lambda(\wx,x)/\partial x_k,
\end{equation}
\begin{equation}\label{dLagr:wp}
\widetilde{p}_k=\partial\Lambda(\wx,x)/\partial \widetilde{x}_k.
\end{equation}

\setcounter{equation}{0}
\section{Simplest flow of the Toda hierarchy\newline
and its bi--Hamiltonian structure}
The both lattices (\ref{new}) and (\ref{Toda}) arise from the simplest flow of 
the Toda hierarchy under two different parametrizations of the 
relevant variables $(a,b)$ (called Flaschka variables) by the canonically 
conjugated variables $(x,p)$.

The simplest flow of the Toda hierarchy is:
\begin{equation}\label{TL}
\dot{a}_k=a_k(b_{k+1}-b_k), \quad \dot{b}_k=a_k-a_{k-1}.
\end{equation}
It may be considered either under open--end boundary conditions
($a_0=a_N=0$), or under periodic ones (all the subscripts are
taken (mod $N$), so that $a_0\equiv a_N$, $b_{N+1}\equiv b_1$). 

It is easy to see that the flow (\ref{TL}) is Hamiltonian with respect to two 
different compatible Poisson brackets. The first of them is linear:
\begin{equation}\label{l br}
\{a_k,b_k\}_1=-\{a_k,b_{k+1}\}_1=a_k
\end{equation}
(only the non--vanishing brackets are written down), and a Hamiltonian function
generating the flow (\ref{TL}) in this bracket is equal to
\begin{equation}\label{H 1}
H^{(1)}=\frac{1}{2}\sum_{k=1}^Nb_k^2+\sum_{k=1}^Na_k.
\end{equation}
The second Poisson bracket is given by:
\begin{equation}\label{q br}
\{b_{k+1},b_k\}_2=a_k, \quad \{a_{k+1},a_k\}_2=a_{k+1}a_k, \quad
\{b_k,a_k\}_2=-b_ka_k, \quad \{b_{k+1},a_k\}_2=b_{k+1}a_k,
\end{equation}
the corresponding Hamiltonian function being
\begin{equation}\label{H 2}
H^{(2)}=\sum_{k=1}^N b_k.
\end{equation}

An integrable discretization of the flow (\ref{TL}) is given by the difference 
equations \cite{PGR}, \cite{S3}
\begin{equation}\label{dTL}
\wa_k=a_k\frac{\beta_{k+1}}{\beta_k},\quad
\wb_k=b_k+h\left(\frac{a_k}{\beta_k}-\frac{a_{k-1}}{\beta_{k-1}}\right), 
\end{equation}
where $\beta_k=\beta_k(a,b)$ is defined as a unique set of functions 
satisfying the recurrent relation
\begin{equation}\label{recur}
\beta_k=1+hb_k-\frac{h^2a_{k-1}}{\beta_{k-1}}
\end{equation}
together with an asymptotic relation
\begin{equation}\label{as beta}
\beta_k=1+hb_k+O(h^2).
\end{equation}
In the open--end case, due to $a_0=0$, we obtain from (\ref{recur}) the 
following finite continued fractions expressions for $\beta_k$:
\[
\beta_1=1+hb_1;\quad 
\beta_2=1+hb_2-\frac{h^2a_1}{1+hb_1};\quad\ldots\quad;
\]
\[
\beta_N=1+hb_N-\frac{h^2a_{N-1}}{1+hb_{N-1}-
\displaystyle\frac{h^2a_{N-2}}{1+hb_{N-2}-
\parbox[t]{1.0cm}{$\begin{array}{c}\\  \ddots\end{array}$}
\parbox[t]{2.2cm}{$\begin{array}{c}
 \\  \\-\displaystyle\frac{h^2a_1}{1+hb_1}\end{array}$}}}.
\]
In the periodic case  (\ref{recur}), (\ref{as beta}) uniquely define 
$\beta_k$'s as $N$-periodic infinite continued fractions. It can be 
proved that for $h$ small enough these continued fractions converge and their 
values satisfy (\ref{as beta}).

It can be proved \cite{S3} that the map (\ref{dTL}) is Poisson with respect 
to the both brackets (\ref{l br}) and (\ref{q br}), and hence with respect
to their arbitrary linear combination.

Let us recall also the Lax representations of the flow (\ref{TL}) and of the
map (\ref{dTL}). They are given in terms of the  $N\times N$ Lax matrix $T$
depending on the phase space coordinates 
$a_k, b_k$ and (in the periodic case) on the additional parameter $\lambda$:
\begin{equation}\label{T}
T(a,b,\lambda) = \sum_{k=1}^N b_kE_{kk}+\lambda\sum_{k=1}^N E_{k+1,k}+
\lambda^{-1}\sum_{k=1}^N a_kE_{k,k+1}.
\end{equation}
Here $E_{jk}$ stands for the matrix whose only nonzero entry on the intersection
of the $j$th row and the $k$th column is equal to 1. In the periodic case we
have $E_{N+1,N}=E_{1,N}, E_{N,N+1}=E_{N,1}$; in the open--end case we set
$\lambda=1$, and $E_{N+1,N}=E_{N,N+1}=0$. 

The flow (\ref{TL}) is equivalent to
the following matrix differential equation:
\begin{equation}\label{Lax}
\dot{T}=\left[ T,B\right], 
\end{equation}
where
\begin{equation}\label{B}
B(a,b,\lambda) = \sum_{k=1}^Nb_kE_{kk}+\lambda\sum_{k=1}^N E_{k+1,k},
\end{equation}
and the map (\ref{dTL}) is equivalent to the following matrix difference
equation:
\begin{equation}\label{dLax}
\widetilde{T}={\rm\bf B}^{-1}T{\rm\bf B},
\end{equation}
where
\begin{equation}\label{bB}
{\rm\bf B}(a,b,\lambda)=\sum_{k=1}^N\beta_kE_{kk}
+h\lambda\sum_{k=1}^NE_{k+1,k}.
\end{equation}

The spectral invariants of the matrix $T(a,b,\lambda)$ serve as 
integrals of motion for the flow (\ref{TL}), as well as for the map
(\ref{dTL}). 

In particular, it is easy to see that the Hamiltonian functions (\ref{H 1}),
(\ref{H 2}) are spectral invariants of the Lax matrix:
\[
H^{(1)}=\frac{1}{2}{\rm tr}(T^2),\quad H^{(2)}={\rm tr}(T).
\]

\setcounter{equation}{0}
\section{Reminding the Toda lattice case}

The Toda lattice (\ref{Toda}) admits a Lagrangian formulation with a 
Lagrange function
\begin{equation}\label{Toda Lagr}
{\cal L}^{(1)}(x,\dot{x})
=\frac{1}{2}\sum_{k=1}^N\dot{x}_k^2-\sum_{k=1}^N\exp(x_k-x_{k-1}).
\end{equation} 
A general procedure implies that the momenta $p_k$ are given by
\[
p_k=\partial {\cal L}^{(1)}/\partial\dot{x}_k=\dot{x}_k,
\]
so that the corresponding Hamiltonian function is
\begin{equation}\label{H 1 in xp}
H^{(1)}=\frac{1}{2}\sum_{k=1}^Np_k^2+\sum_{k=1}^N\exp(x_k-x_{k-1}),
\end{equation}
and the flow (\ref{TL}) takes the form of canonical equations of motion:
\begin{eqnarray*}
\dot{x}_k & = & \partial H^{(1)}/\partial p_k=p_k,\\
\dot{p}_k & = & -\partial H^{(1)}/\partial x_k=
\exp(x_{k+1}-x_k)-\exp(x_k-x_{k-1}).
\end{eqnarray*}

One sees immediately that this coincides with the flow (\ref{TL}), if
the Flaschka variables $(a,b)$ are introduced according to the formulas
\begin{equation}\label{l par}
a_k=\exp(x_{k+1}-x_k),\quad b_k=p_k.
\end{equation}
Obviously, this leads immediately to the linear Poisson brackets (\ref{l br}).

Let us turn now to the discrete time case. Consider first the equations of
motion (\ref{Sur2}). It is easy to see that they admit a Lagrangian formulation 
with the Lagrange function
\begin{equation}\label{Sur2:Lagr}
\Lambda_1(\wx,x)=\sum_{k=1}^N\phi_1(\wx_k-x_k)-h\sum_{k=1}^N\exp(x_k-\wx_{k-1}),
\end{equation}
where  $\phi_1(\xi)=(\exp(\xi)-1-\xi)/h$. Hence they are equivalent to the
symplectic map $(x,p)\mapsto(\wx,\wip)$ with
\begin{equation}\label{Sur2:p}
hp_k = \exp(\wx_k-x_k)-1+h^2\exp(x_k-\wx_{k-1}),
\end{equation}
\begin{equation}\label{Sur2:wp}
h\wip_k = \exp(\wx_k-x_k)-1+h^2\exp(x_{k+1}-\wx_k).
\end{equation}
We demonstrate now that they may be put in the form (\ref{dTL}).

{\bf Proposition 1.} {\it If the variables $a_k$, $b_k$ are defined by}
(\ref{l par}), {\it and
\begin{equation}\label{Sur2:beta}
\beta_k=\exp(\wx_k-x_k),
\end{equation}
then} (\ref{Sur2:p}), (\ref{Sur2:wp}) {\it imply} (\ref{dTL}), (\ref{recur}).

{\bf Proof.} The first equation of motion in (\ref{dTL}) follows immediately
from the definitions of $a_k=\exp(x_{k+1}-x_k)$, $\beta_k=\exp(\wx_k-x_k)$.
The recurrent relation (\ref{recur}) is just a reformulation of (\ref{Sur2:p})
in the variables $a_k$, $b_k$, $\beta_k$. Finally, the second equation in 
(\ref{dTL}) follows immediately from (\ref{Sur2:wp}) and (\ref{Sur2:p}). \qed

Note that this proposition implies immediately that the map (\ref{dTL})
is Poisson with respect to the linear bracket (\ref{l br}).

A very remarkable circumstance was found in \cite{S3}: an apparently different 
discretization (\ref{Sur3}) is in fact only another parametrization of the
same map (\ref{dTL}). It is easy to see that (\ref{Sur3}) admits a Lagrangian
formulation with a Lagrange function
\begin{equation}\label{Sur3:Lagr}
\Lambda_2(\wx,x)=\sum_{k=1}^N\frac{1}{2h}(\wx_k-x_k)^2-
\sum_{k=1}^{N}\phi_2(x_k-\wx_{k-1}),
\end{equation}
where $\phi_2(\xi)=h^{-1}\int_0^\xi\log(1+h^2\exp(\eta))d\eta$.
(It is easy to see that the corresponding equations (\ref{d Lagr}) read:
\[
\wx_k-2x_k+\undertilde{x_k}=
\log\Big(1+h^2\exp(\undertilde{x_{k+1}}-x_k)\Big)-
\log\Big(1+h^2\exp(x_k-\wx_{k-1})\Big),
\]
which is equivalent to (\ref{Sur3})). Hence an equivalent form of writing
(\ref{Sur3}) in canonically conjugated variables $(x,p)$ is:
\begin{equation}\label{Sur3:p}
\exp(hp_k)=\exp(\wx_k-x_k)\Big(1+h^2\exp(x_k-\wx_{k-1})\Big)=
\exp(\wx_k-x_k)+h^2\exp(\wx_k-\wx_{k-1}),
\end{equation}
\begin{equation}\label{Sur3:wp}
\exp(h\wip_k)=\exp(\wx_k-x_k)\Big(1+h^2\exp(x_{k+1}-\wx_k)\Big)=
\exp(\wx_k-x_k)+h^2\exp(x_{k+1}-x_k).
\end{equation}

{\bf Proposition 2.} {\it If the variables $a_k$, $b_k$ are defined by}
\begin{equation}\label{mod l par}
a_k=\exp(x_{k+1}-x_k+hp_k),\quad 1+hb_k=\exp(hp_k)+h^2\exp(x_k-x_{k-1}),
\end{equation}
{\it and
\begin{equation}\label{Sur3:beta}
\beta_k=\exp(hp_k),
\end{equation}
then} (\ref{Sur3:p}), (\ref{Sur3:wp}) {\it imply} (\ref{dTL}), (\ref{recur}).

{\bf Proof.} From (\ref{Sur3:p}), (\ref{Sur3:wp}), and the first equation in
(\ref{mod l par}) it follows that the first equation of motion in (\ref{dTL})
is satisfied, if
\[
\beta_k=\exp(\wx_k-x_k)\Big(1+h^2\exp(x_k-\wx_{k-1})\Big),
\]
which is just (\ref{Sur3:beta}). Now (\ref{recur}) is a reformulation of the
second equation in (\ref{mod l par}), if one takes into account that
$a_k/\beta_k=\exp(x_{k+1}-x_k)$. The second equation of motion in (\ref{dTL})
follows directly from the second equation in (\ref{mod l par}), (\ref{Sur3:wp}), 
and (\ref{Sur3:p}). \qed

It is easy to calculate that a Poisson bracket for the variables $a_k$, $b_k$ 
resulting from the parametrization (\ref{mod l par}) reads:
\begin{equation}
\begin{array}{c}
\{b_{k+1},b_k\}=ha_k,\quad \{a_{k+1},a_k\}=ha_{k+1}a_k,\\ \\
\{b_{k+1},a_k\}=a_k+hb_{k+1}a_k,\quad \{b_k,a_k\}=-a_k-hb_ka_k,
\end{array}
\end{equation}
which is exactly a linear combination $\{\cdot,\cdot\}_1+h\{\cdot,\cdot\}_2$.

We conclude this section by noting that the both Lagrange functions
(\ref{Sur2:Lagr}) and (\ref{Sur3:Lagr}) serve as difference approximations
to the continuous time one (\ref{Toda Lagr}).

\setcounter{equation}{0}
\section{A new lattice}

We turn now to the system (\ref{new}). First of all, on sees readily that 
it admits a Lagrangian formulation with
\begin{equation}\label{new Lagr}
{\cal L}^{(2)}(x,\dot{x})=\sum_{k=1}^N[\dot{x}_k\log(\dot{x}_k)-\dot{x}_k]-
\sum_{k=1}^N\exp(x_k-x_{k-1}).
\end{equation}
Hence the momenta $p_k$ are introduced by
\[
p_k=\partial{\cal L}^{(2)}/\partial\dot{x}_k=\log(\dot{x}_k),
\]
hence the corresponding Hamiltonian function is equal to
\begin{equation}\label{H 2 in xp}
H^{(2)}=\sum_{k=1}^N \exp(p_k)+\sum_{k=1}^N\exp(x_k-x_{k-1}),
\end{equation}
and the canonical form of the equations of motion is:
\begin{eqnarray*}
\dot{x}_k &=&\partial H^{(2)}/\partial p_k=\exp(p_k),\\
\dot{p}_k &=&-\partial H^{(2)}/\partial x_k=
\exp(x_{k+1}-x_k)-\exp(x_k-x_{k-1}).
\end{eqnarray*}

It is now easy to see that if one introduces variables $a_k$, $b_k$ 
according to
\begin{equation}\label{q par}
a_k=\exp(x_{k+1}-x_k+p_k), \quad b_k=\exp(p_k)+\exp(x_k-x_{k-1}),
\end{equation}
then their evolution induced by the flow above just coincides with
(\ref{TL}).

It can be readily checked that (\ref{q par}) leads to Poisson brackets
(\ref{q br}), and that the notation $H^{(2)}$ for the function
(\ref{H 2 in xp}) is consistent with (\ref{H 2}).

So, the equations (\ref{new}) admit a Lax representation (\ref{Lax})
with the matrices (\ref{T}), (\ref{B}), for the entries of which one has the
formulas (\ref{q par}), which is equivalent also to
\[
a_k=\dot{x}_k\exp(x_{k+1}-x_k),\quad b_k=\dot{x}_k+\exp(x_k-x_{k-1}).
\]

Turning to the discrete time system (\ref{dTL}), we find the following
results. It admits a Lagrangian formulation with
\begin{equation}\label{dnew1 Lagr}
\Lambda_3(\wx,x)=\sum_{k=1}^N\phi(\wx_k-x_k)-
\sum_{k=1}^{N}\psi(x_k-\wx_{k-1}),
\end{equation}
where the two functions $\phi(\xi), \psi(\xi)$ are defined by
\begin{equation}\label{phipsi}
\phi(\xi)=\int_0^{\xi}\log\left|\frac{\exp(\eta)-1}{h}\right|d\eta,\quad
\psi(\xi)=\int_0^{\xi}\log(1+h\exp(\eta))d\eta.
\end{equation}
Hence a symplectic map $(x,p)\mapsto(\wx,\wip)$ generated by (\ref{dnew1}) 
may be defined by the following relations:
\begin{eqnarray}
h\exp(p_k) & = & \Big(\exp(\wx_k-x_k)-1\Big)\Big(1+h\exp(x_k-\wx_{k-1})\Big),
\label{dnew1:p}\\
h\exp(\wip_k) & = &
\Big(\exp(\wx_k-x_k)-1\Big)\Big(1+h\exp(x_{k+1}-\wx_k)\Big).\label{dnew1:wp}.
\end{eqnarray}
It is very remarkable that this map can be again reduced to (\ref{dTL})!

{\bf Proposition 3.} {\it If the variables $a_k$, $b_k$ are defined by}
(\ref{q par}), {\it and
\begin{equation}\label{dnew1:beta}
\beta_k=\exp(\wx_k-x_k)\Big(1+h\exp(x_k-\wx_{k-1})\Big),
\end{equation}
then} (\ref{dnew1:p}), (\ref{dnew1:wp}) {\it imply} (\ref{dTL}), (\ref{recur}).

{\bf Proof.}  The first equation of motion in (\ref{dTL}) follows immediately
from $a_k=\exp(x_{k+1}-x_k+p_k)$ and (\ref{dnew1:p}), (\ref{dnew1:wp}),
(\ref{dnew1:beta}). The recurrent relation (\ref{recur}) follows from
(\ref{dnew1:beta}), (\ref{dnew1:p}), if one takes into account that
\begin{equation}\label{dnew1:aux}
h^2a_k/\beta_k=h\Big(\exp(x_{k+1}-x_k)-\exp(x_{k+1}-\wx_k)\Big),
\end{equation}
and hence
\[
1+h\exp(x_k-x_{k-1})-\frac{h^2a_{k-1}}{\beta_{k-1}}=1+h\exp(x_k-\wx_{k-1}).
\]
The second equation of motion follows from (\ref{q par}), (\ref{dnew1:p}), 
and (\ref{dnew1:wp}) with the help of (\ref{dnew1:aux}). \qed

\setcounter{equation}{0}
\section{Discretizations related to \newline
relativistic Toda lattice}

We now turn to the discretizations (\ref{Hir}), (\ref{Sur1}). 
A simple observation shows that these models are equivalent to  
(\ref{Sur2}), (\ref{Sur3}), respectively, when considered as equations on 
the lattice with the coordinates $(t,k)$. More precisely, the equations of 
motion (\ref{Sur2}), (\ref{Sur3}) are recovered 
from (\ref{Hir}), (\ref{Sur1}) after renaming $x_k(t)$ to $x_k(t-kh)$. However, 
such renaming mixes the "spatial" and "temporal" variables, and this changes the
properties of the {\it initial value problem}, which we are concerned 
with, dramatically.

First of all, from a practical point of view we must 
remark that the Hirota's and the standard--like models are explicit 
with respect to $\wx_k$, while the models (\ref{Sur2}), (\ref{Sur3})
require to solve certain nonlinear algebraic equations (or, equivalently,
to evaluate continued fractions) in order to obtain the $\wx_k$.

Another important difference between our new models and the old ones 
lies in their algebraic, $r$--matrix structure. According to the 
observation in \cite{PGR}, the Hirota's and the standard--like models are in 
essence equivalent. More precisely, they both arise from the following
system of difference equations:
\begin{equation}\label{dRTL}
\wid_k+h^2\wc_{k-1}=d_k+h^2c_k,\quad \wid_{k+1}c_k=d_k\wc_k,
\end{equation}
if the variables $(c,d)$ are parametrized by canonically conjugated variables
$(x,p)$ in two different ways. An equivalent form of equations (\ref{dRTL})
may be obtained, if one resolves for $(\wc_k,\wid_k)$:
\begin{equation}\label{res dRTL}
\wid_k=d_{k-1}\frac{d_k+h^2c_k}{d_{k-1}+h^2c_{k-1}},\quad
\wc_k=c_k\frac{d_{k+1}+h^2c_{k+1}}{d_k+h^2c_k}.
\end{equation}

The map defined by these difference equations is Poisson with respect to two
different compatible Poisson brackets: a linear one,
\begin{equation}\label{r l br}
\{c_k,d_{k+1}\}_1=-c_k, \quad \{c_k,d_k\}_1=c_k,\quad \{d_k,d_{k+1}\}_1=h^2c_k,
\end{equation}
and a quadratic one,
\begin{equation}\label{r q br}
\{c_k,c_{k+1}\}_2=-c_kc_{k+1}, \quad \{c_k,d_{k+1}\}_2=-c_kd_{k+1}, \quad
\{c_k,d_k\}_2=c_kd_k.
\end{equation}

The Lax representation for the map (\ref{dRTL}) may be given in terms of the
$N\times N$ matrices depending on the dynamical variables $(c,d)$ and an
additional parameter $\lambda$:
\begin{eqnarray}
L(c,d,\lambda) & = & \sum_{k=1}^N d_kE_{kk}+h\lambda\sum_{k=1}^N E_{k+1,k},\\
U(c,d,\lambda) & = & \sum_{k=1}^N E_{kk}-h\lambda^{-1}\sum_{k=1}^N 
c_kE_{k,k+1}.
\end{eqnarray}
It is easy to check that the difference equations (\ref{dRTL}) are 
equivalent to the matrix equation
\begin{equation}\label{LU=UL}
U\widetilde{L}=L\widetilde{U},\quad{\rm or}\quad 
\widetilde{L}\widetilde{U}^{-1}=U^{-1}L.
\end{equation}
In terms of the Lax matrix 
\begin{equation}\label{r T}
T(c,d,\lambda)=L(c,d,\lambda)U^{-1}(c,d,\lambda)
\end{equation}
the equation (\ref{LU=UL}) takes the form
\begin{equation}
\widetilde{T}=U^{-1}TU=L^{-1}TL,
\end{equation}
which implies, in particular, that the spectral invariants of the matrix $T$
are integrals of motion for the map (\ref{dRTL}).

As observed in \cite{S1}, \cite{S2}, the matrix $T$ from (\ref{r T}) just coincides
with the Lax matrix of the {\it relativistic Toda hierarchy} (which is also
bi--Hamiltonian with respect to both brackets (\ref{r l br}), (\ref{r q br})).

We first recall how can the equations (\ref{Hir}), (\ref{Sur1}) be reduced 
to (\ref{dRTL}), and then show that the same is true for (\ref{dnew2}).

We start with (\ref{Hir}). It is easy to find a Lagrangian formulation of these
equations with a Lagrange function
\begin{equation}\label{Hir:Lagr}
\Lambda_4(\wx,x)=\sum_{k=1}^N\phi_1(\wx_k-x_k)-
h\sum_{k=1}^N\exp(\wx_k-\wx_{k-1}),
\end{equation}
(where, as in the previous section,  $\phi_1(\xi)=(\exp(\xi)-1-\xi)/h$). Hence 
the equations (\ref{Hir}) are equivalent to a symplectic map 
$(x,p)\mapsto(\wx,\wip)$ with
\begin{eqnarray}
hp_k & = & \exp(\wx_k-x_k)-1,\label{Hir:p}\\
h\wip_k & = & \exp(\wx_k-x_k)-1+
h^2\exp(\wx_{k+1}-\wx_k)-h^2\exp(\wx_k-\wx_{k-1}).\label{Hir:wp}
\end{eqnarray}

{\bf Proposition 4.} {\it Let the coordinates $(c,d)$ be parametrized by the
canonically conjugated varibles $(x,p)$ according to the formulas
\begin{equation}\label{r l par}
c_k=\exp(x_{k+1}-x_k),\quad d_k=1+hp_k-h^2\exp(x_{k+1}-x_k).
\end{equation}
Then} (\ref{Hir:p}), (\ref{Hir:wp}) {\it imply} (\ref{dRTL}).

{\bf Proof.} Obviously, we have from (\ref{Hir:p}), (\ref{Hir:wp}), and
(\ref{r l par}):
\[
d_k+h^2c_k=\exp(\wx_k-x_k),\qquad \wid_k+h^2\wc_{k-1}=\exp(\wx_k-x_k).
\]
Comparing these expressions, we get the first equation of motion in
(\ref{dRTL}), and the first of the expressions above together with
$c_k=\exp(x_{k+1}-x_k)$ implies the second equation in (\ref{res dRTL}).
\qed

It is important to notice that the parametrization (\ref{r l par}) results 
in the linear Poisson bracket (\ref{r l br}), which proves independently
that the map (\ref{res dRTL}) is Poisson with respect to this bracket.

Turning now to (\ref{Sur1}), we find a Lagrangian formulation of these equations
(in the form (\ref{stand})) with
\begin{equation}\label{Sur1:Lagr}
\Lambda_5(\wx,x)=\sum_{k=1}^N\frac{1}{2h}(\wx_k-x_k)^2-
\sum_{k=1}^{N}\phi_2(x_k-x_{k-1}),
\end{equation}
where, as in the previous section, $\phi_2(\xi)=
h^{-1}\int_0^\xi\log(1+h^2\exp(\eta))d\eta$.
Hence the expression for the momenta $p_k$ and their updates, equivalent to
(\ref{Sur1}), are:
\begin{eqnarray}
\exp(hp_k) & = &
\exp(\wx_k-x_k)\;\frac{1+h^2\exp(x_k-x_{k-1})}{1+h^2\exp(x_{k+1}-x_k)},
\label{Sur1:p}\\
\exp(h\wip_k) & = & \exp(\wx_k-x_k),\label{Sur1:wp}
\end{eqnarray}

{\bf Proposition 5.} {\it Let the coordinates $(c,d)$ be parametrized by the
canonically conjugated varibles $(x,p)$ according to the formulas
\begin{equation}\label{r q par}
c_k=\exp(x_{k+1}-x_k+hp_k), \quad d_k=\exp(hp_k).
\end{equation}
Then} (\ref{Sur1:p}), (\ref{Sur2:wp}) {\it imply} (\ref{dRTL}).

{\bf Proof.} It is easy to see that (\ref{r q par}) allows to rewrite the
second equation in (\ref{dRTL}) as $\exp(x_{k+1}-x_k+h\wip_{k+1})=
\exp(\wx_{k+1}-\wx_k+h\wip_k)$. This equality is an obvious consequence of
(\ref{Sur1:wp}). Further, (\ref{r q par}) allows to rewrite the first equation 
in (\ref{res dRTL}) as
\[
\exp(\wip_k)=\exp(p_k)\frac{1+h^2\exp(x_{k+1}-x_k)}{1+h^2\exp(x_k-x_{k-1})},
\]
which follows immediately from (\ref{Sur1:p}), (\ref{Sur2:wp}). \qed

This time we notice that (\ref{r q par}) results (up to the factor $h$) 
in the quadratic Poisson bracket (\ref{r q br}), which proves independently
that the map (\ref{res dRTL}) is Poisson with respect to this bracket.

It remains to perform analogous considerations for an explicit discretization
(\ref{dnew2}) of our new lattice (\ref{new}). Remarkably, this system turns
out to be still another realization of the same map (\ref{res dRTL})!
To demonstrate this, note that (\ref{dnew2}) admits a Lagrangian formulation
with
\begin{equation}\label{dnew2 Lagr}
\Lambda_6(\wx,x)=\sum_{k=1}^N\phi(\wx_k-x_k)-
\sum_{k=1}^{N}\psi(x_k-x_{k-1}),
\end{equation}
where $\phi(\xi), \psi(\xi)$ are defined by (\ref{phipsi}).
Hence a Hamiltonian formulation of this system is given by:
\begin{eqnarray}
h\exp(p_k) & = &
\Big(\exp(\wx_k-x_k)-1\Big)\;\frac{1+h\exp(x_k-x_{k-1})}{1+h\exp(x_{k+1}-x_k)},
\label{dnew2:p}\\
h\exp(\wip_k) & = & \Big(\exp(\wx_k-x_k)-1\Big),\label{dnew2:wp}
\end{eqnarray}

{\bf Proposition 6.} {\it Let the coordinates $(c,d)$ be parametrized by the
canonically conjugated varibles $(x,p)$ according to the formulas
\begin{equation}\label{new par}
c_k=\exp(x_{k+1}-x_k+p_k),\quad d_k=1+h\exp(p_k)+h\exp(x_k-x_{k-1}).
\end{equation}
Then} (\ref{dnew2:p}), (\ref{dnew2:wp}) {\it imply} (\ref{dRTL}).

{\bf Proof.} From (\ref{new par}) and (\ref{dnew2:p}) it follows:
\begin{eqnarray}
d_k+h^2c_k & = & 1+h\exp(x_k-x_{k-1})+h\exp(p_k)(1+h\exp(x_{k+1}-x_k))
\nonumber\\
           & = & \exp(\wx_k-x_k)\Big(1+h\exp(x_k-x_{k-1})\Big).
\label{dnew2:aux1}
\end{eqnarray}
Analogously, from (\ref{new par}) and (\ref{dnew2:wp}) it follows:
\begin{eqnarray}
\wid_k+h^2\wc_{k-1} & = & 1+h\exp(\wip_k)+h\exp(\wx_k-\wx_{k-1})
(1+h\exp(\wip_{k-1}))\nonumber\\
 & = & \exp(\wx_k-x_k)\Big(1+h\exp(x_k-x_{k-1})\Big).\label{dnew2:aux2}
\end{eqnarray}
Comparing these expressions, we get the first equation of motion in (\ref{dRTL}).
The second equation in (\ref{res dRTL}) is a direct consequence of 
(\ref{dnew2:p}), (\ref{dnew2:wp}), and (\ref{dnew2:aux1}). \qed

It is easy to calculate that the parametrization (\ref{new par}) 
generates the following Poisson bracket:
\begin{equation}
\begin{array}{c}
\{c_{k+1},c_k\}=c_{k+1}c_k,\quad \{d_{k+1},d_k\}=h^2c_k,\\ \\
\{d_k,c_k\}=c_k-d_kc_k,\quad \{d_{k+1},c_k\}=-c_k+d_{k+1}c_k.
\end{array}
\end{equation}
This is, obviously, a linear combination of the brackets (\ref{r l br})
and (\ref{r q br}), namely $\{\cdot,\cdot\}_2-\{\cdot,\cdot\}_1$. Of course,
the Poisson property of the map (\ref{res dRTL}) with respect to this
bracket follows from the previous results, but the Proposition 6 gives
an alternative way to prove this.

\setcounter{equation}{0}
\section{Conclusion}
Identifying the variables $(a,b)$ in (\ref{l par}) and in (\ref{q par}), we
get a transformation between two sets of variables $(x,p)$ (and, consequently,
between two sets of variables $(\dot{x},x)$). This is exactly
the B\"acklund transformation between the lattice (\ref{new}) and the Toda 
lattice (\ref{Toda}).

For each of these systems one has different integrable discretizations. Some
of them share the Lax matrix with the continuous time prototype. These
discretizations generate Newtonian equations implicit with respect to the
updates $\wx_k$. Other discretizations have Lax representations with the
Lax matrix defining the {\it relativistic} Toda hierarchy. These discretizations
turn out to be explicit. All three apparently different implicit discretizations
turn out to be connected by B\"acklund transformations. An underlying fact is
that all three appear from one and the same integrable map, if the relevant
variables $(a,b)$ are parametrized by canonically conjugated ones $(x,p)$  in
three different ways, generating three different Poisson brackets on the 
set of $(a,b)$ (and hence on the set of Lax matrices). Exactly the same holds
true for the three explicit discretizations.

We would like to note here that all the Poisson brackets on the sets of Lax
matrices were given an $r$--matrix interpretation in \cite{S2}.


\begin{thebibliography}{10}
\bibitem[1]{H} R.Hirota. Nonlinear partial difference equations.
II. Discrete--time Toda equation. -- {\it J. Phys. Soc. Japan, 1977,
v.43, p.2074--2078;}\\
IV. B\"acklund transformations for the discrete--time Toda equation. -- 
{\it J. Phys. Soc. Japan, 1978, v.45, p.321--332.}

\bibitem[2]{S1} Yu.B.Suris. Generalized Toda chains in discrete time. -- 
{\it Leningrad Math. J., 1991, v.2,  p.339--352;}\\
Discrete--time generalized Toda lattices: complete integrability and relation 
with relativistic Toda lattices. -- {\it Physics Letters A, 1990, v.145,
p.113--119;}\\
Algebraic structure of discrete--time and relativistic Toda 
lattices. -- {\it Physics Letters A, 1991, v.156, p.467--474.}

\bibitem[3]{S2} Yu.B.Suris. On the bi--Hamiltonian structure of Toda and relativistic  Toda 
lattices. -- {\it Physics Letters A, 1993, v.180, p.419--429.}

\bibitem[4]{PGR} V.Papageorgiou, B.Grammaticos, A.Ramani. Orthogonal polynomial
approach to discrete Lax pairs for initial--boundary value problems of the $QD$
algorithm. -- To appear in: {\it Letters Math. Phys, 1995, v.34.}

\bibitem[5]{S3} Yu.B.Suris. Bi--Hamiltonian structure of the $qd$ algorithm
and new discretizations of the Toda lattice . -- 
{\it Physics Letters A, 1995, v.206, p.153--161}.
\end{thebibliography}
\end{document}